\documentclass{aa}
\usepackage{graphicx}

\begin{document}

\title{The role of magnetic helicity transport in nonlinear galactic
dynamos}
\author{N.\, Kleeorin \inst{1} \and D.\, Moss \inst{2} \and I.\,
Rogachevskii\inst{1}  \and D.\, Sokoloff\inst{3}} \offprints{I.
Rogachevskii} \institute{Department of Mechanical Engineering,
Ben-Gurion
University of Negev, POB 653,  84105 Beer-Sheva, Israel\\
\email{nat@menix.bgu.ac.il; gary@menix.bgu.ac.il}
\and
Department of Mathematics, University of Manchester, Manchester M13
9PL, UK \\
\email{moss@maths.man.ac.uk}
\and Department of Physics, Moscow State University, Moscow
119899, Russia \\
\email{sokoloff@dds.srcc.msu.su}}

\date{Received: 8 January 2002; accepted: 8 March, 2002}

\abstract{We consider the magnetic helicity balance for the
galactic dynamo in the framework of the local dynamo problem, as
well as in the no-$z$ model (which includes explicitly the radial
distribution of the magnetic fields).  When calculating the
magnetic helicity balance we take into account the redistribution
of the small-scale and large-scale magnetic fields between the
magnetic helicities,  as well as magnetic helicity transport and
diffusion due to small-scale turbulence. We demonstrate that the
magnetic helicity flux through the galactic disc boundaries leads
to a steady-state magnetic field with magnetic energy comparable
to the equipartition energy of the turbulent motions of the
interstellar medium. If such flux is ignored, the steady-state
magnetic field is found to be  much smaller than the equipartition
field. The total magnetic helicity flux through the boundaries
consists of both an advective flux and a diffusive flux. The exact
ratio of these contributions seems not to be crucial for
determining the strength of the steady-state magnetic field and
its structure.  However at least some diffusive contribution is
needed to smooth the magnetic helicity profile near to the disc
boundaries. The roles of various transport coefficients for
magnetic helicity are investigated, and the values which lead to
magnetic field configurations comparable with those observed are
determined.

\keywords{Galaxies: magnetic fields}}

\maketitle

\section{Introduction}

The magnetic fields of spiral galaxies are thought to be generated
by hydromagnetic dynamo action, based on the joint action of the
mean helicity of the interstellar  turbulence and the differential
rotation (see e.g.,  Ruzmaikin et al. 1988). The growth of a
dynamo generated magnetic field must eventually be constrained by
some nonlinear effect. The most simple and straightforward idea
concerning the nonlinear constraint is based on equipartition
between the kinetic energy of interstellar turbulence and the
magnetic energy. More precisely, the $\alpha$-effect based on the
hydrodynamic helicity is presumed to be quenched by equipartition
strength magnetic fields. Although the idea is very crude, this
assumption results in models of galactic magnetic fields which are
quite similar to the observed magnetic field distributions in
spiral galaxies (see Beck et al. 1996, for a review).

This viewpoint was strongly criticized in the  early 1990s by
Vainshtein \& Cattaneo (1992) and Gruzinov \& Diamond (1995), who
suggested that the dynamo growth of the large-scale galactic
magnetic field will saturate when its energy is lower than the
equipartition value by many orders of magnitudes. As it has been
recognized quite recently (Kleeorin et al. 2000; Blackman \& Field
2000), their arguments are based implicitly on local conservation
of magnetic helicity: as magnetic helicity is a conserved
quantity, it can only change if there is a flux of magnetic
helicity through the boundaries, or through resistive effects
which are, however, very slow.  On the other hand, dynamo models
predict growth of the magnetic helicity of the large-scale
magnetic field.  Since the total magnetic helicity is locally
conserved, the small-scale and large-scale contributions to the
magnetic helicity must have opposite signs. Vainshtein \&
Cattaneo's conclusions  follow, since the small-scale magnetic
field must possess magnetic helicity of equal magnitude (but of
opposite sign). This field saturates the $\alpha$-effect and hence
dynamo action, long before equipartition of the large-scale field
is attained.

Several possibilities for overcoming the problem have been suggested,
and new insight into the physical mechanisms of nonlinear saturation
supports the applicability of conventional galactic magnetic field
models (for a review, see Kulsrud, 1999).  In particular, Moss et al.
(1999) stress that magnetic helicity conservation arguments do not
take into account additional magnetic helicity generation arising
from the buoyancy of the large scale magnetic field, as suggested by
Parker (1992), and showed that galactic dynamo models based on this
mechanism yield  magnetic field configurations that are quite similar
to those observed.

Moreover, Kleeorin et al. (2000) stress that galactic dynamo
action is impossible without a turbulent flux of magnetic field
through the boundaries of the galactic disc (cf. Zeldovich et al.
1983) and it is quite reasonable to suggest that magnetic helicity
is also transported through the disc boundaries. Kleeorin \&
Rogachevskii (1999) proposed a quantitative model for this flux of
magnetic helicity and Kleeorin et al. (2000) demonstrated that
dynamo action is saturated at the equipartition level of the
large-scale field, provided that the magnetic helicity flux is
taken into account.

It remains however unclear to what extent the magnetic helicity
flux through the boundaries can contribute to the resolution of
the magnetic helicity conservation problem (see e.g.  Brandenburg
et al. 2001a). In particular, saturation processes involve many
other effects in addition to any flux of magnetic helicity through
boundaries. It is necessary to make a careful distinction between
the helicities and the $\alpha$-effect, and to consider nonlinear
quenching of turbulent magnetic diffusivity, etc. A more specific
point is that Kleeorin et al. (2000) considered a quite crude
galactic dynamo model, replacing magnetic helicity losses by a
simple decay term and considering magnetic helicity saturation as
a process that occurs much faster than the dynamo growth.  The aim
of the present paper is to remove some of these restrictions and,
to some extent, to clarify these questions.

We first consider a consistent, nonlinear, one dimensional model
for the local galactic dynamo generation, based on arguments of
magnetic helicity conservation, including turbulent magnetic
helicity transport through the boundaries, magnetic helicity
diffusion due to small-scale turbulence and nonlinear
$\alpha$-quenching, and distinguish between helicities and the
$\alpha$-effect. Modern theoretical studies provide specific forms
for the effects under discussion. However, we take into account
that the forms used are, to an extent, uncertain, and could be
model-dependent, and so we consider our models also in a more
phenomenological light, by varying values of parameters by an
order of magnitude. We obtain a quantitative agreement between our
model and the observed distribution of galactic magnetic field.
Unsurprisingly perhaps, the `best' values of the model parameters
differ from the theoretical predictions.  Note that we do not here
take into account any nonlinear quenching of the turbulent
magnetic diffusion, which is the subject of a separate study. We
believe that the effects of these nonlinearities are less
important than those modifying the $\alpha$-effect, and so choose
here to isolate the latter. A quantitative model for a nonlinear
quenching of turbulent magnetic diffusion has been recently
suggested by Rogachevskii \& Kleeorin (2001).

Note that Brandenburg et al. (2001b) also obtained equipartition
field strengths in their 3D simulations with periodic boundary
conditions. However, the time scales were rather large for high
magnetic Reynolds numbers. The main reason is that the full
treatment of the back reaction of the magnetic field onto the
turbulence also involves a turbulent diffusivity quenching. Such a
quenching is beyond the scope of the present paper, and so we
cannot expect to reproduce this result of Brandenburg et al.
(2001b) at the moment by using our approach. However there is no
obstacle  in principle in including a quenching of the turbulent
diffusion in future studies.

Our result is not completely straightforward  regarding estimation
of the role of the flux through the boundaries (cf. Brandenburg et
al. 2001a). We find that all the effects taken into account
participate in the saturation process at a comparable level.
Moreover, our model includes two kinds of magnetic helicity flux,
i.e. explicit magnetic helicity transport as well as diffusion of
magnetic helicity through the boundaries due to small-scale
turbulence. The action of both effects is quite similar, and it is
difficult to distinguish between them from modelling of the
steady-state magnetic field distribution.

We also perform 2D simulations of the  galactic dynamo in the
framework of the so-called "no-$z$" model. This involves
observable properties only, in that it ignores the unobservable
vertical magnetic field profiles in galactic discs, but retains
the horizontal field structure. The steady-state field
distributions obtained are quite similar to those known from
observations, and from models with a naive $\alpha$-quenching.

\section{The equations for the magnetic and current helicities}

The equation for the current helicity $\langle \vec{b} \cdot
(\vec{\nabla} {\bf \times} \vec{b}) \rangle$ can be obtained using
arguments based on the magnetic helicity conservation law. We
introduce the large-scale vector potential $\vec{A}$, small-scale
vector potential $\vec{a}$, and the corresponding representations
for the magnetic fields, $\vec{B}$ and $\vec{b}$. We then write
the total magnetic field as $\vec{H} = \vec{B} + \vec{b}$, and the
total vector potential as $\vec{\cal A}=\vec{A}+\vec{a}$, so that
the fields are decomposed into mean and fluctuating parts.  The
equation for the vector potential $\vec{\cal A}$ follows from the
induction equation for the total magnetic field $\vec{H}$

\begin{eqnarray}
\partial \vec{\cal A}
/ \partial t = \vec{v}{\bf \times}\vec{H}  - \eta \, \vec{\nabla}{\bf
\times}\vec{H} +  \vec{\nabla} \varphi \;,
\label{B2}
\end{eqnarray}
where $ \vec{v} = \vec{V} + \vec{u} ,$ and $ \vec{V} = \langle
\vec{v} \rangle $ is the mean fluid velocity field, $ \eta $ is the
magnetic diffusion due to the electrical conductivity of the fluid
and $ \varphi $ is an arbitrary scalar function. Now we multiply the
induction equation for the total magnetic field $\vec{H}$ by
$\vec{a}$ and Eq. (\ref{B2}) by $\vec{b}$, add them and average over
the ensemble of turbulent fields. This yields an equation for the
magnetic helicity $ \chi^{m} = \langle \vec{a} \cdot \vec{b} \rangle
$ in the form

\begin{eqnarray}
\partial \chi^{m} / \partial t +
\vec{\nabla} \cdot \vec{F} = - 2 \langle \vec{u}{\bf \times}\vec{b}
\rangle \cdot \vec{B} - 2 \eta \langle \vec{b} \cdot (\vec{\nabla}
{\bf \times} \vec{b}) \rangle \;, \label{B3} \end{eqnarray} where
$\vec{F} = (2/3) \vec{V} \chi^{m} + \langle \vec{a} {\bf \times}
(\vec{u}{\bf \times} \vec{B}) \rangle - \eta \langle \vec{a} {\bf
\times} (\vec{\nabla} {\bf \times} \vec{b}) \rangle + \langle \vec{a}
{\bf \times} (\vec{u} {\bf \times} \vec{b}) \rangle - \langle \vec{b}
\varphi \rangle $ is the flux of magnetic helicity. The electromotive
force for an isotropic and homogeneous turbulence is

\begin{eqnarray}
\langle \vec{u}{\bf \times}\vec{b} \rangle = \alpha \vec{B} -
\eta_{T} \, (\vec{\nabla} {\bf \times} \vec{B}) \;, \label{B4}
\end{eqnarray}
where $ \eta_{T} $ is the turbulent magnetic diffusivity, and it
is assumed that $ \alpha $ is the total alpha-effect which at the
nonlinear stage includes both the original hydrodynamic and the
magnetic contributions.  Now we take into account that $ \chi^{c}
\equiv \tau \langle \vec{b} \cdot (\vec{\nabla} {\bf \times}
\vec{b}) \rangle $ is proportional to the magnetic helicity:
$\chi^{c} = \chi^{m} / (18 \pi \eta_{T} \rho)$ (see, e.g.,
Kleeorin \& Rogachevskii, 1999), where $ \rho $ is the density and
$ \tau $ is the correlation time of turbulent velocity field. We
would like to stress that $ \chi^{c} $ is a modified current
helicity, i.e. the current helicity multiplied by the correlation
time $ \tau $ of velocity field. Thus, the equation for $ \chi^{c}
$ in dimensionless form is given by

\begin{equation}
{{\partial \chi^{c}} \over {\partial t}} + {\chi^{c} \over T} +
\vec{\nabla} \cdot \vec{F} = 4 (h/l)^2 [R_\alpha^{-1} \, \vec{B}
\cdot (\vec{\nabla} {\bf \times} \vec{B}) - \alpha B^2] \; \label{B5}
\end{equation}
(see Kleeorin \& Ruzmaikin, 1982; Kleeorin \& Rogachevskii, 1999).
We adopt here the standard dimensionless form of the galactic
dynamo equations from Ruzmaikin et al. (1988); in particular,
length is measured in units of the disc thickness $h$, time in
units of $ h^{2} / \eta_{T} $ and $B$ is measured in units of the
equipartition energy $B_{\rm eq} = \sqrt{4 \pi \rho} \, u_\ast $.
$\chi^c$ and $ \alpha $ are in units of $ \alpha_\ast $ (the
maximum value of the hydrodynamic part of the $ \alpha $ effect).
We define $R_\alpha = h \alpha_\ast / \eta_{T} ,$ $ T = (1/3)
(l/h)^{2} Rm $, where $l$ is the maximum scale of the turbulent
motions, $ Rm = l u_{\ast} / \eta $ is the magnetic Reynolds
number. Also $u_{\ast}$ is the characteristic turbulent velocity
at the scale $l$, $\rho$ is the gas density, and the turbulent
diffusivity $\eta_T=lu_{\ast}/3$.  For galaxies the term $\chi^{c}
/ T$ is very small and can be dropped.

The turbulent flux $\vec F$ of magnetic helicity is proportional
to the hydrodynamic part of the $\alpha$ effect and the turbulent
magnetic diffusivity (see Kleeorin \& Rogachevskii, 1999; Kleeorin
et al. 2000). The turbulent flux of magnetic helicity serves as an
additional nonlinear source in the equation for $ \chi^{c} $ and
it causes a drastic change in the dynamics of the large-scale
magnetic field.  Eq. (\ref{B5}) is the basic equation used below
to describe the nonlinear evolution of the galactic dynamo. We
adopt below a specific form of the flux $\vec F$ which includes a
transport of the magnetic helicity, leading to an advected  flux
through the boundaries, as well as a diffusion flux due to the
small-scale turbulence.

\section{Basic equations}
\label{basic}

The standard mean field dynamo equation takes the well-known form

\begin{equation}
\label{mfd} \frac{\partial {\vec{B}}}{\partial t}=
\vec{\nabla}\times(\vec{V}\times
\vec{B}+\alpha\vec{B}-\eta_T\vec{\nabla}\times\vec{B}).
\end{equation}
The analysis leading to the standard local thin disc dynamo equations
for an axisymmetric magnetic field can be found in, for example,
Ruzmaikin et al.  (1988).  Using cylindrical polar coordinates $r,
\phi, z$, we obtain the equations for the mean radial field
$B_r=R_\alpha b_r$ and toroidal field $B_\phi$ for the local
thin-disc $\alpha \Omega$-dynamo problem as

\begin{eqnarray}
{{\partial b_r} \over {\partial t}} &=& - (\alpha (\vec{B}) B_\phi)'
+ b_r'' \;, \label{A1} \\ {{\partial B_\phi} \over {\partial t}} &=&
D b_r + B''_\phi \;.
\label{A2}
\end{eqnarray}
Here $ b' = \partial
b / \partial z ,$ $D=r \, (d\Omega/dr) \, h^2/\eta_T$ is the dynamo
number, and $\alpha (\vec{B})$ is the total $\alpha$ effect which is
given by

\begin{eqnarray}
\alpha(\vec{B}) = \alpha^v + \alpha^m \;,
\label{AA3}
\end{eqnarray}
where $\alpha^v$ is the hydrodynamic part of the $\alpha$ effect, and
$\alpha^m$ is the magnetic part of the $\alpha$ effect.
These quantities are determined by the corresponding
helicities and quenching functions, $\phi_{v}(B)$ and $\phi_{m}(B) ,$
and $B=|\vec{B}|.$ In particular, $\alpha^v = \chi^v \phi_{v}(B) ,$
$\alpha^m = \chi^c(\vec{B}) \phi_{m}(B) $ and $\chi^v = - (\tau /3)
\langle \vec{u} \cdot (\vec{\nabla} {\bf \times} \vec{u}) \rangle .$
Thus,

\begin{eqnarray} \alpha(\vec{B}) =
\chi^v \phi_{v}(B) + \chi^c(\vec{B}) \phi_{m}(B) \; , \label{A3}
\end{eqnarray}
where the quenching functions $\phi_{v}(B)$ and $\phi_{m}(B)$ are
given by

\begin{eqnarray}
\phi_{v}(B) &=& (4/7) \phi_{m} (B) +
(3/7)[1 - 16 B^{2} \nonumber \\
&+& 128 B^{4} \ln (1 + (8B^{2})^{-1})] \;,
\label{A4} \\
\phi_{m} (B) &=& (3 / 8 B^{2}) (1 -\arctan (\sqrt{8} B) / \sqrt{8} B) \;   \label{A5}
\end{eqnarray}
(see Rogachevskii \& Kleeorin, 2000).  Thus $\phi_{v}(B) =
1/(4B^2)$ and $\phi_{m}(B) = 3/(8B^2)$ for $B \gg 1/\sqrt{8};$ and
$\phi_{v}(B) = 1-(48/5)B^2$ and $\phi_{m}(B) = 1-(24/5)B^2$ for $B
\ll 1/\sqrt{8}.$

The function $\chi^c(\vec{B})$ entering the magnetic part  of the
$\alpha$ effect is determined by the evolutionary equation

\begin{eqnarray}
{\partial \chi^c \over \partial t} &=& 4 \biggl({h
\over l} \biggr)^2 [B_\phi b'_r - B'_\phi b_r - \alpha(\vec{B}) B^2
\nonumber \\
&+& C{{\partial} \over {\partial z}} (|\chi^v(z)|
\phi_{v}(B) B^2)] + \kappa (\chi^c)'' \; ,
 \label{A6}
\end{eqnarray}
where we have written $ \vec{B} \cdot (\vec{\nabla} {\bf \times}
\vec{B}) = R_\alpha (B_\phi b'_r - b_r B'_\phi) ,$ and $ b' =
\partial b / \partial z$, and have introduced a coefficient $C$ for the
magnetic helicity flux transport term, for future convenience.
Unless otherwise stated, $C=1$. We have also introduced an
explicit diffusion of $\chi^c$, with coefficient $\kappa$.

This diffusion plays an important role below; however it does not
follow directly from the arguments of magnetic helicity
conservation presented in the previous section. A consistent way
to introduce it into the equation of magnetic helicity balance
requires a more detailed analysis of the interstellar turbulence
at the nonlinear stage of the galactic dynamo. We consider here
that the random flows present in the interstellar medium consist
of a combination of small-scale motions, which are affected by
magnetic forces resulting in a steady-state of the dynamo, and a
microturbulence which is supported by a strong random driver
(supernovae explosions) which can be considered as independent of
the galactic magnetic field. The large-scale magnetic field is
smoothed over both kinds of turbulent fluctuations, while the
small-scale magnetic field is smoothed over microturbulent
fluctuations only. It is the smoothing over the microturbulent
fluctuations that gives the coefficient $\kappa$. In particular,
the relevant analysis can be performed using Eqs.~(9)--(11) in
Kleeorin \& Rogachevskii (1999). More pragmatically, it is
difficult to imagine that a turbulent diffusion acts on the
magnetic field, but not on the current helicity.

Equations (\ref{A3})--(\ref{A6}) contain the main nonlinearities
suggested for the nonlinear $ \alpha $ effect. For example, the
function $\phi_{v}(B)$ describes conventional quenching of the $
\alpha $ effect. A simple form of such a quenching, $\phi_v = 1/(1
+ B^{2}) ,$ was introduced long ago (see, e.g., Iroshnikov, 1970;
R\"udiger, 1974; Roberts \& Soward, 1975). The splitting of the
total $\alpha$ effect into the hydrodynamic, $\alpha^v ,$ and
magnetic, $\alpha^m ,$ parts was first suggested by Frisch et al.
(1975).  The magnetic part $\alpha^m$ includes two types of
nonlinearity: the algebraic quenching described by the function
$\phi_{m}(B)$ (see, e.g., Field et al., 1999; Rogachevskii and
Kleeorin, 2000) and the dynamic nonlinearity which is determined
by Eq. (\ref{A6}).  This equation describes the evolution of
magnetic helicity, i.e.  its production, dissipation and
transport.  The governing equation for magnetic helicity was
proposed by Kleeorin \& Ruzmaikin (1982; see also the discussion
by Zeldovich et al., 1983) for an isotropic turbulence, and
investigated by Kleeorin et al. (1995) for stellar dynamos, and
self-consistently derived by Kleeorin \& Rogachevskii (1999) for
an arbitrary anisotropic turbulence. Schmalz \& Stix (1991) and
Covas et al. (1998) also investigated related solar dynamo models.
Magnetic helicity transport through the boundary of a dynamo
region is reported by Chae (2001) to be observable at the solar
surface.  The equation for $\chi^{c}$ in the form of Eq.
(\ref{A6}) was given by Kleeorin et al. (2000), but without the
term with coefficient $\kappa$. Note that the role of a flux of
magnetic helicity in the dynamics of the mean magnetic field in
accretion discs was recently discussed by Vishniac \& Cho (2001).

\section{The local thin disc model}

\subsection{Asymptotic expansions and  an equilibrium solution}

We now present asymptotic expansions for a galactic dynamo model
determined by Eqs. (\ref{A1})--(\ref{A6}).
In a steady-state, Eq. (\ref{A6}) with $ \kappa = 0 $ gives
\begin{eqnarray} \alpha(\vec{B}) B^{2} = B_\phi b'_r - B'_\phi b_r +
C [|\chi^v(z)| \phi_{v}(B) B^{2}]' \; .  \label{C1} \end{eqnarray}
Then, for fields of even parity with respect to the disc plane, Eqs.
(\ref{A1}) and (\ref{A2}) give
\begin{eqnarray}
B'''_\phi &+& D \alpha (B) B_\phi = 0 \;, \label{C2} \\ b_r &=& -
D^{-1}
B''_\phi \; .
\label{C3}
\end{eqnarray}
For the $\alpha \Omega$-dynamo $ B \approx B_\phi .$ This assumption
is justified if $|D| \gg R_\alpha ,$ i.e. $|R_\omega| \gg 1 .$ Thus
Eqs. (\ref{C1})--(\ref{C3}) yield
\begin{equation}
[(B')^{2}]' - 2 \, C \, D \, (|\chi^v(z)| \phi_{v}(B) B^{2})' = 0 \; .
\label{C4}
\end{equation}
For an arbitrary  profile $|\chi^v(z)|$ and negative
dynamo number $D$, there is an explicit steady solution of this
equation with the boundary conditions $ B_{\phi}(z=1) = 0 $ and $
B_{\phi}'(z=0) = 0 $,
\begin{eqnarray}
B(z) = (C \, |D| / 2)^{1/2}
\int_{|z|}^{1} \sqrt{|\chi^v(\tilde z)|} \,d \tilde z \;,
\label{C5}
\end{eqnarray}
where $z$ is measured in the units of $h ,$ and we
have assumed that $B \gg 1/\sqrt{8} .$ For the specific choice of the
hydrodynamic helicity profile $|\chi^v(z)| =\sin^{2}(\pi z/ 2)$ we
obtain
\begin{eqnarray}
B_\phi &=& {{\sqrt {2 C \, |D|}} \over {\pi}}
B_{\rm eq} \cos \, \biggl({{\pi z} \over 2}\biggr) \;,
\label{C6}\\
B_r &=& - {\pi \sqrt{2 C \, R_\alpha} \over 4 \sqrt {|R_\omega|} }
B_{\rm eq} \cos \, \biggl({{\pi z} \over 2}\biggr) \;,
\label{C7}
\end{eqnarray}
where we have restored the dimensional factor $B_{\rm
eq} .$ The boundary conditions for $B_{r}$ are $B_{r}(z=1) = 0$ and $
B_{r}'(z=0) = 0 .$ The pitch angle of the magnetic field lines is $p=
- {\rm arctan} \, (\pi^2/4|R_\omega|) \approx 14^\circ$ for $|D| =
10$ and $R_\alpha = 1 .$ Note, however, that our asymptotic analysis
in the vicinity of the point $z=1$ is not self-consistent.  The
numerical results (Sect. 4.2) demonstrate that in this
region the role of $\kappa$ is important.  Note also that for profiles
with $\chi^v(z=1) = 0$ a steady-state solution for $B_{r}$ does not
satisfy the boundary condition $B_{r}(z=1) = 0 .$ This is the reason
why we choose the profile with  $\chi^v(z=1) \not= 0 .$

Now we obtain a steady-state solutions for $ C = 0 $ and $ \kappa
\not= 0 .$ Thus, Eq. (\ref{A6}) yields:
\begin{eqnarray}
B_\phi b'_r
- B'_\phi b_r - \alpha(\vec{B}) B^{2} + \tilde \kappa (\chi^c)'' = 0
\;,
\label{CC1}
\end{eqnarray}
where $ \tilde \kappa = \kappa / 4
(h/l)^2 .$ For fields of even parity with respect to the disc plane,
Eqs. (\ref{A3}), (\ref{C2}), (\ref{C3}) and (\ref{CC1}) give for $B
\gg 1/\sqrt{8} $:
\begin{eqnarray}
[(B')^{2} - 2 |D| \tilde \kappa
(\chi^c)']' = 0 \;,
\label{CC4}
\end{eqnarray}
where $ B \approx
B_\phi $ for the $\alpha \Omega$-dynamo.  Combining Eqs. (\ref{CC1})
and (\ref{CC4}) we obtain
\begin{eqnarray}
4 (B B^{(IV)} &+& B' B''') - (3 / 4 \tilde \kappa) (B')^{2} =
- |D| [(\chi^v)'
\nonumber\\
&+& (3/2) (\chi^c)'_{z=0}] \;,
\label{CC2}
\end{eqnarray}
where $ B^{(IV)} $ is
the fourth-order $z$-derivative.  For the specific choice of the
hydrodynamic helicity profile $\chi^{v} = \sin(\pi z)$ there is a
steady-state solution
\begin{eqnarray}
B_\phi &=& {2 \over {\pi}} \biggl({|D| \over \pi^{2} + 3/ (8 \tilde
\kappa)} \biggr)^{1/2} B_{\rm eq} \cos \, \biggl({{\pi z} \over
2}\biggr) \;, \label{CC3}\\ B_r &=& - {{\pi} \over 2}
\biggl({R_{\alpha} \over |R_{\omega}| (\pi^{2} + 3/ (8 \tilde
\kappa))} \biggr)^{1/2} B_{\rm eq} \cos \, \biggl({{\pi z} \over
2}\biggr) \;, \label{CC5} \end{eqnarray} which exists for the
boundary condition $ (\chi^{c})'_{z=0} = - (4 \pi^{2} \tilde \kappa +
3/2)^{-1} .$ Note that the term  proportional to $ \kappa (\chi^c)''
$ in Eq.  (\ref{A6}) can be interpreted as a flux of magnetic
helicity which causes a steady-state solution for the mean magnetic
field. The pitch angle for this solution turns out  to be the same as
for that with $\kappa =0$.

\subsection{Numerical solutions}
\label{num1}

We found solutions of Eqs.~(\ref{A1} - \ref{A6}) by step by step
integration, from arbitrarily chosen initial conditions.

In agreement with our previous results (Kleeorin et al. 2000), the
flux terms connected with nonvanishing coefficients $C$ and $\kappa$
lead to (large-scale) magnetic field evolution at significant
steady-state magnetic field strengths, i.e. at strengths comparable to
the
equipartition value.  However, the relative roles of $C$ (magnetic
helicity transport) and $\kappa$ (magnetic helicity diffusion due to
the small-scale turbulence) appear to be quite unexpected (note that
the analysis of Kleeorin et al., 2000 was too crude to make a
distinction between these effects). We obtained steady-state
solutions even for vanishing $C$ and nonvanishing $\kappa$, while a
weak  diffusion of magnetic helicity is needed in order to get a
steady-state with nonvanishing $C$. This fact should be connected
with the peculiar behavior of the asymptotic solution near to the
disc boundaries when $\kappa =0$, mentioned in Sect. 4.1.
However this distinction between the role of magnetic helicity
transport and diffusion seems to be more of a mathematical rather
than a physical nature. For $\kappa = 0$, Eq. (\ref{A6}) is a first
order equation with respect to the spatial variables, whereas it is a
second order equation for $\kappa \ne 0$, and this difference causes
problems
with the boundary conditions. Presumably, it means that the
evaluation of the term proportional to $C$ in this equation, whilst
self-consistent inside the galactic disc, needs some improvement near
to the sharp boundary. In general, we find that both effects destroy
the local
conservation of magnetic helicity and that their variations affect the
final magnetic field distribution in a similar way.

We used spatial profiles $\chi^v=\sin^2(\pi z/2)$, $\chi^v=\sin
(\pi z)$, $\chi^v=\sin^2(\pi z)$, for values of the ratio $h/l=2,
5, 10$, and put $C=1, 10$. The solutions are steady, except for
$D=-1000$ and $h/l\geq 5$, in the cases with either
$\chi^v=\sin(\pi z)$ or $\chi^v= \sin^2(\pi z/2)$, and $C=10$.

Various properties of these solutions are illustrated in Figs.~1 -- 5.
It can be seen clearly that the field strength ($\sim |B_\phi|$) is
typically of order 1 (equipartition) or larger, and increases with
$|D|$.  The role of the flux transport term in Eq.~(\ref{A6}) can be
seen in Fig.~3, i.e., the magnitude of the mean magnetic field
$|\vec{B}|$ increases with the coefficient $C$.

\begin{figure}
\centering
\includegraphics[width=8cm]{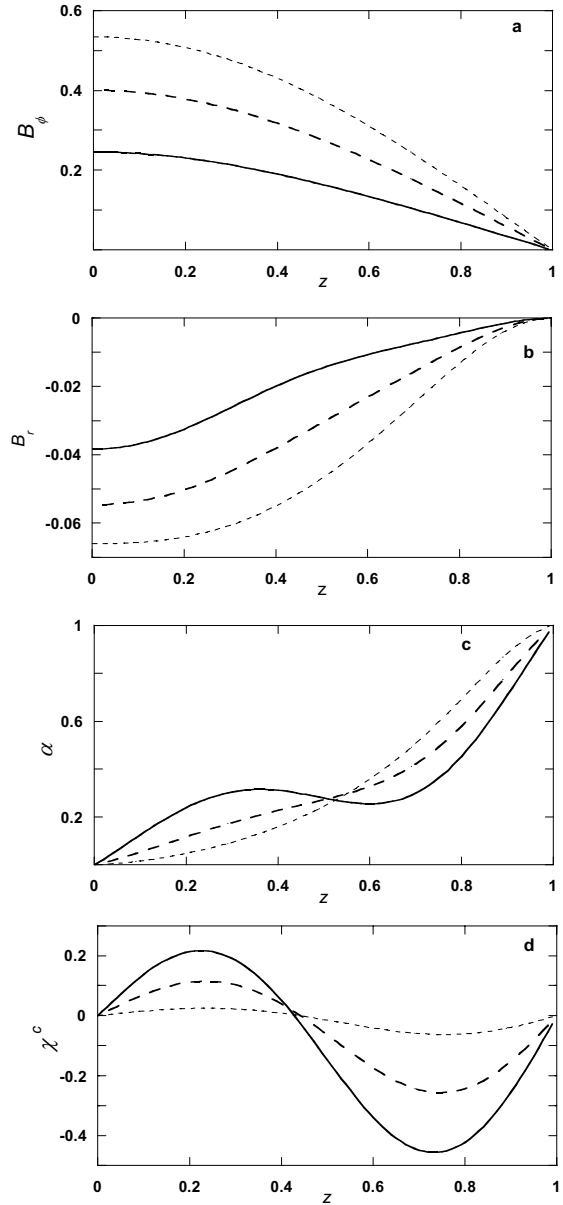}
\caption{The $z$-dependence of solutions for the one-dimensional model; dynamo
number $D=-20$, $C=1$, $\kappa=1$
and $|\chi^v(z)|=\sin^2(\pi z/2)$.
The various curves indicate results with $h/l=2$
(short-dashed), $h/l=5$ (dashed) and $h/l=10$ (solid).
a) the toroidal magnetic field $B_{\phi}(z)$, b) the poloidal magnetic
field $B_{r}(z)$, c) the total $\alpha$-effect, $\alpha(z)$,
d) the function $\chi^{c}(z)$.}
\label{Fig1.ps}
\end{figure}

\begin{figure}
\centering
\includegraphics[width=8cm]{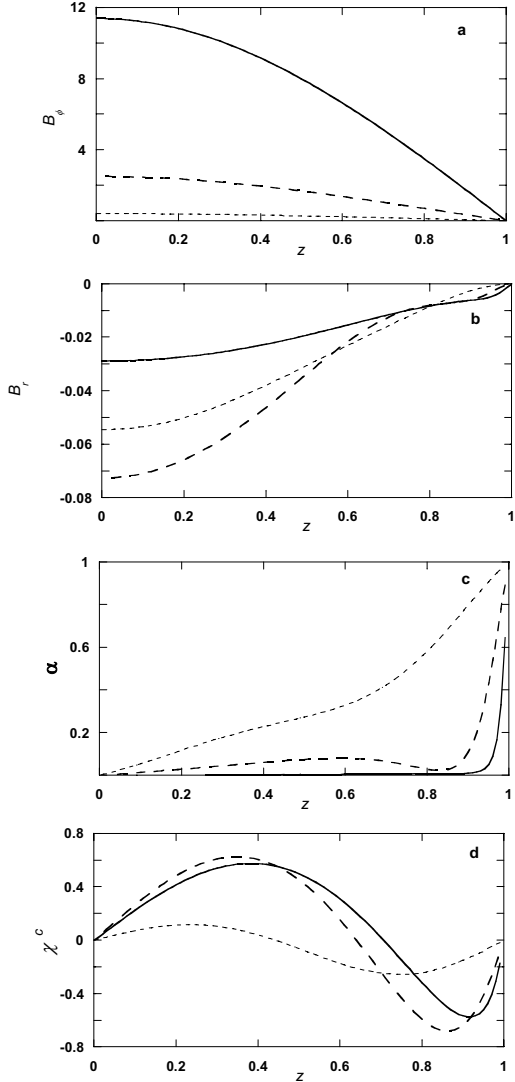}
\caption{As Fig.~\protect\ref{Fig1.ps}, with $h/l=5$, and dynamo numbers
$D=-20$ (short-dashed), $D=-100$ (dashed) and
$D=-1000$ (solid).}
\label{Fig2.ps}
\end{figure}

\begin{figure}
\centering
\includegraphics[width=8cm]{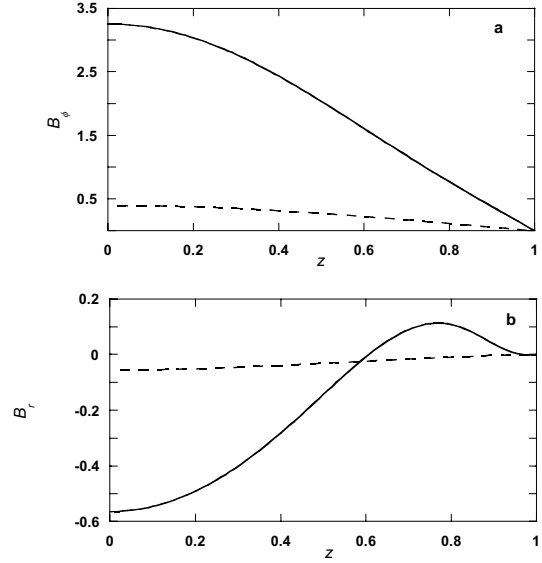}
\caption{As Fig.~\protect\ref{Fig1.ps}, with $D=-20$,
showing the effects of changing $C$:
$C=1$ (dashed) and $C=10$ (solid); $h/l = 5$.}
\label{Fig3.ps}
\end{figure}

\begin{figure}
\centering
\includegraphics[width=7cm]{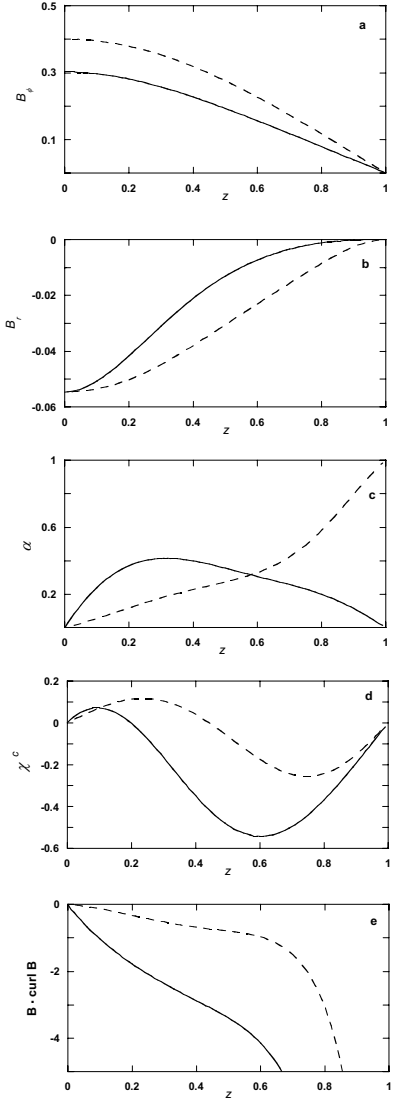}
\caption{As Fig.~\protect\ref{Fig1.ps}, with $D=-20$, $C=1$, $h/l=5$, showing
the effects of changing the form of $\chi^v$:
$|\chi^{v}(z)| =
\sin^{2}(\pi z/2)$ (dashed), and $\chi^{v}(z) = \sin(\pi z)$ (solid).
}
\label{Fig4.ps}
\end{figure}

\begin{figure}
\centering
\includegraphics[width=7cm]{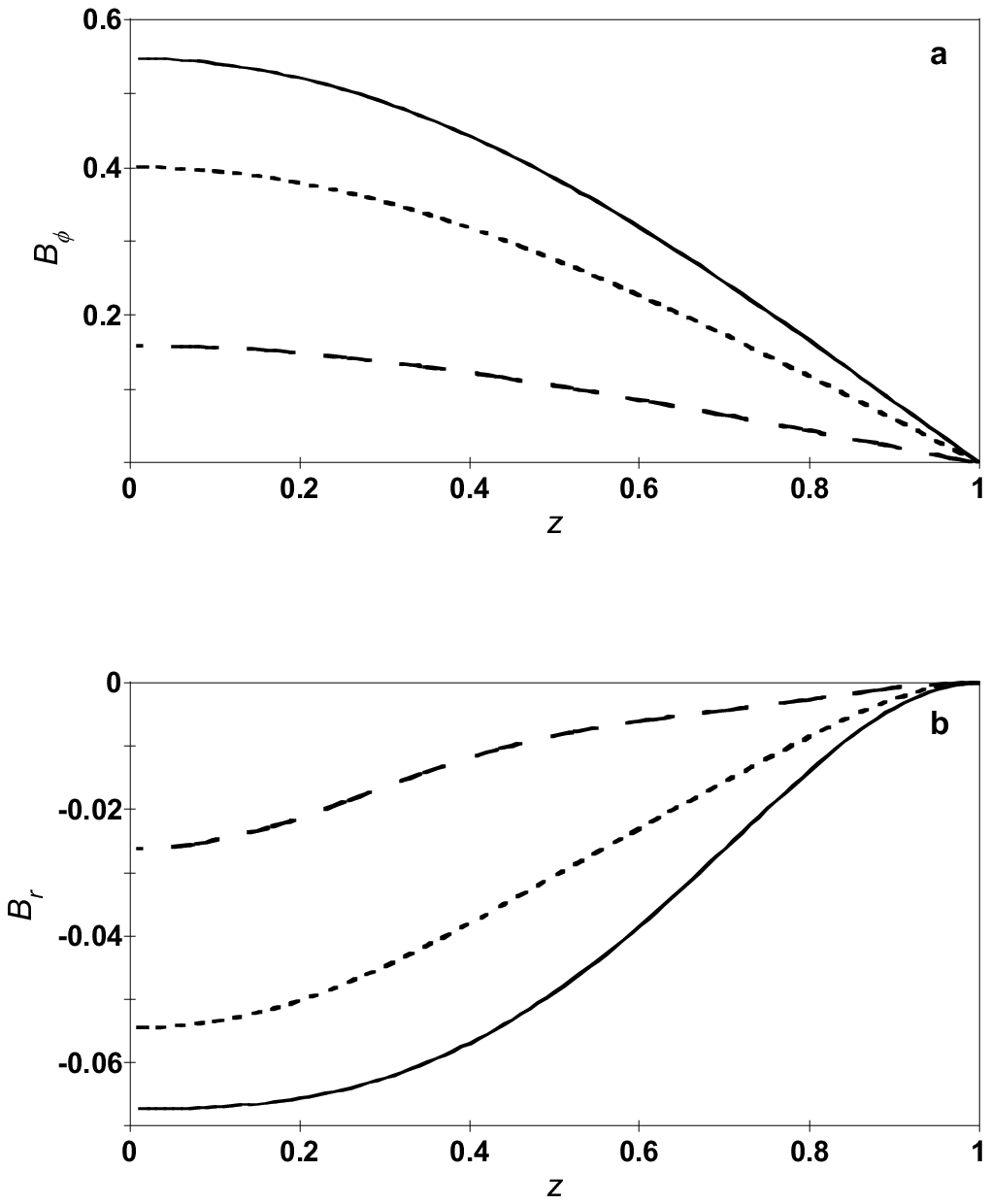}
\caption{As Fig.~\protect\ref{Fig1.ps}, with $D=-20$, $C=1$, $h/l=5$,
$|\chi^{v}(z)| = \sin^{2}(\pi z/2)$, showing the effects of changing $\kappa$:
$\kappa=0.1$ (dashed), $\kappa=1$
(short-dashed), $\kappa=10$ (solid). The values of $\chi^c$ vary in
the ranges $(-0.5, 0.25)$, $(-0.25, 0.10)$ and $(-0.04, 0.15)$ for
$\kappa = 0.1, \, 1$ and 10 respectively. } \label{Fig5.ps}
\end{figure}

However, it is clear  that a detailed agreement with observed
galactic magnetic fields needs some effort. In many cases we obtain a
steady-state magnetic field about $0.1 - 0.5 B_{\rm eq}$ while
observations suggest  $B \approx B_{\rm eq}$. To get this we need
$|D| > 20 - 100$ (Fig.~2), which is a quite restrictive condition, or
a value of $C$ that exceeds unity (Fig. 3).  However straightforward
theoretical estimations suggest $C<1$. Even so, the values of $C$
required do not appear unreasonable, given the inherent uncertainties
of the theory.

\section{The no-$z$ model}

\subsection{The model equations}

The basic no-$z$ dynamo model for disc galaxies is described in Moss
(1995).
Phillips (2001) suggested a plausible tuning of the model and his
major amendment, the multiplication by factors $\pi^2/4$ of the terms
representing the $z$-diffusion of $B_r$ and $B_\phi$, was implemented
here.  The no-$z$ model differs from the local model of
Sect.~\ref{basic} in that it describes magnetic fields over the
entire radial range, $0\leq r\leq R$, but all explicit dependence on
the vertical coordinate $z$ has been removed, with $z$-derivatives
being replaced by inverse powers of $h$.  The field components $B_r$,
$B_\phi$ appearing in the no-$z$ equations can either be thought of
as representing mid-plane values, or as some sort of vertical
average of values
through the disc.  Here we develop a version of Eq. (\ref{A6}) for
the no-$z$ model, starting from the
appropriate analogue of Eq. (\ref{A6}) which is
\begin{eqnarray}
{{\partial \chi^c} \over {\partial t}} &=& \rho(r)^{-1} \biggl\{ 4
\biggl({h \over l} \biggr)^2 \biggl[R_\alpha^{-1} \vec{B} \, \cdot
(\vec{\nabla} \times\vec{B}) - \alpha B^2
\nonumber\\
&& - {\rm div} \biggl({\vec{\nabla} \rho \over \rho} \, \chi^v \,B^2
\Phi_v(B/\sqrt\rho) \biggr) \biggr]
\nonumber\\
&& + \Delta \chi^c - {\rm div}[\vec{V} \, \chi^c] \biggr \} \;,
\label{D1} \\
\alpha &=& \chi^v \, \Phi_v (B/\sqrt \rho) + \chi^c \,
\Phi_m (B/\sqrt \rho) \;,
\label{D2}
\end{eqnarray}
where $\vec{V} = \vec{e}_{\phi} \, \Omega \, r $ is the
large-scale velocity (differential rotation). The factor
$\rho^{-1}$ multiplying by the right hand side of Eq. (\ref{D1})
arises because the link between the magnetic helicity $\chi^m$ and
the current helicity $\chi^c$ contains the density $\rho ,$ i.e.
$\chi^c = \chi^m /18 \pi \eta_T \rho .$ The quenching functions
$\Phi_v$ and  $\Phi_{m}$ contain in their arguments the factor
$1/\sqrt{\rho}$ because they are based on local equipartition,
while Eqs. (\ref{D1}) and (\ref{D2}) presume that the field $B$ is
measured in units of equipartition at the point $r =0$. Note that
our model includes a flux of magnetic helicity which is not
directly connected with a mean magnetic energy flux; however an
energy flux associated with the small-scale magnetic field is, of
course, possible.

The coordinate version of Eq. (\ref{D1}) is given by
\begin{eqnarray}
{\partial \chi^c \over \partial t} &=& \rho^{-1} \biggl \{ 4
\biggl({h \over l} \biggr)^2 [R_\alpha^{-1} (B_\phi B_r' - B_r
B'_\phi) - \alpha B^2
\nonumber\\
&& + (|\chi^v| \, \Phi_v \, B^2)']
+ {4 \lambda^2 \over r} {\partial \over \partial r} [r \, \chi^v \,
\Lambda_\rho^{-1} \, \Phi_v \, B^2] + (\chi^c)''
\nonumber\\
&& + {\lambda^{2} \over r} {\partial \over \partial r} \biggl(r {\partial
\chi^c \over \partial r} \biggr) + {\lambda^2 \over r^{2}}
{\partial ^2 \chi^c \over \partial \phi^2} - \lambda^2 \Omega
{\partial \chi^c \over \partial \phi}  \biggr \} \;,
\label{D3}
\end{eqnarray}
where $(\chi^c)' = \partial \chi^c / \partial z $,
$\lambda= h/R$ is the aspect ratio, and $\Lambda^{-1}_\rho = -
|\vec{\nabla}| \rho / \rho $ is the density stratification scale.

For the no-z model in the axisymmetric case Eq. (\ref{D3})
becomes
\begin{eqnarray}
{\partial \chi^c \over \partial t} &=& \rho^{-1} \biggl \{ 4
\biggl({h \over l} \biggr)^2 [R_\alpha^{-1}\zeta B_r B_\phi + C \,
\chi^v \, \Phi_v(B/\sqrt \rho) \, B^2
\nonumber\\
&& - \alpha B^2] - \kappa \chi^c
+ {4\lambda^2  \over r} {\partial \over \partial r} [
r \, \Lambda_{\rho}^{-1} \, \chi^v \, \Phi_v(B/\sqrt \rho) \, B^2]
\nonumber\\
&& + {\lambda^2 \over r} {\partial \over \partial r} \biggl(r \, {\partial
\chi^c \over \partial r} \biggr) \biggr\}  \; .
\label{D4}
\end{eqnarray}

Note that with  the no-$z$ formalism, the term $\vec{B} \, \cdot
(\vec{\nabla} \times\vec{B})$ vanishes identically. Examination of
the solutions of Sect.~\ref{num1} suggests that setting this term
to zero may not always be a good approximation, but
experimentation with the inclusion of an order of magnitude
estimate for the omitted term suggests that the no-$z$ solutions
are insensitive to it.  We have introduced a coefficient $\kappa$
in the term $-\kappa \, \chi^c$ in Eq.~(\ref{D4}).  The motivation
is that this term represents the term $\partial^2\chi^c/\partial
z^2$ in Eq.~(\ref{D3}). The literal application of the no-$z$
rules would imply $\kappa=1$. However the results of
Sect.~\ref{num1} suggest that in the local model $\chi^c \sim
\sin(2\pi z)$, giving $\kappa \approx 4\pi^2$. This is analogous
to the modification of the representation of the $z$-diffusion
terms for the magnetic field suggested by Phillips (2001), and
mentioned above.  Thus we introduce $\kappa$ as a free parameter.
The term $\chi^v \, \Phi_v(B/\sqrt{\rho}) \, B^2$ represents a
magnetic helicity flux.  As it is introduced in  a somewhat {\it
ad hoc} manner, we have introduced the coefficient $C$ to
represent this uncertainty.  To determine the magnetic field
distribution along the radius we use a Brandt rotation law,
$\Omega(r) = \Omega_0 /[1 + (r/r_\omega)^2]^{1/2}$ with $r_\omega
= 0.2 ,$ and the radial density profile $\rho(r) =
\exp[-(r/r_\rho)^2]$ with $r_\rho = 0.5$, so that $\Lambda_\rho =
2r/r_\rho^2 .$ We also set $\chi^v(r) =1.$

One technical point should be noted here, which is important when
comparing results from the local thin-disc model, studied in the
earlier parts of this paper, with those from the no-$z$ model. For
the local thin-disc model, $R_\omega=r h^2 \eta_T^{-1}
d\Omega/dr<0$.  By the nature of the model, $rd\Omega/dr$ is the
value at a chosen radius in the disc, and $r$ does not further
occur explicitly in the analysis. However the no-$z$ model is {\it
global} with respect to radius, and the value of $rd\Omega /dr$
varies through the disc, from zero at $r=0$ to some maximum
absolute value; for the Brandt rotation law this value is
$0.31\Omega_0$ at $r=\sqrt{2/3}r_\omega$. For the no-$z$ model the
global definition is $R_\omega=R_\omega({\rm
no}-z)=\Omega_0h^2/\eta_T>0$.  (Less importantly, there are also
small differences, of order 25\%, in the effective values of
$R_\alpha$ occurring in the two approximations, even though the
formal definitions are the same - see Phillips (2001).) To
emphasize these distinctions, we write the no-$z$ dynamo number as
${\cal D}=R_\omega({\rm no}-z)R_\alpha$.

For these reasons, it cannot be expected that there will be a precise
correspondence between the values of the marginal dynamo numbers for
these two approximations; empirically we found ${\cal D}_{\rm
crit}\approx 20$.  We note that $|D_{\rm crit}|/{\cal D}_{\rm
crit}\approx 0.4$, which can be compared with the maximum value of
$|rd\Omega /dr|/\Omega_0 = 0.31$ for the no-$z$ model with our
Brandt rotation law.  These issues are also touched on in Sects. 3.2
and 4 of Moss et al. (2001).

What seem to be important are the terms with $\lambda^2 .$ They
represent a diffusion and an advection of helicity
along the radius, i.e. from the galactic center to its periphery.
This could lead to a more extended magnetic field distribution than
given by conventional models. This is why the result can be compared
with that from the conventional model with $\alpha \sim 1/(1 + B^2/
\rho) .$

\subsection{Numerical results}
\label{num2}
We integrated the standard axisymmetric no-$z$ equations (modified as
described above)
for $B_r$ and $B_\phi$, together with Eq.~(\ref{D4}),
from arbitrarily chosen initial conditions for a range of values of
$C$ and $\kappa$, finding steady solutions.
We present in Fig.~6, as  examples  which are close enough
to the available phenomenology of galactic magnetic fields, solutions
with $\kappa=1$, $\zeta =0$, ${\cal D} = 50$, $C=3$ (dashed line) and
$C=10$ (solid line).

\begin{figure}
\centering
\includegraphics[width=7cm]{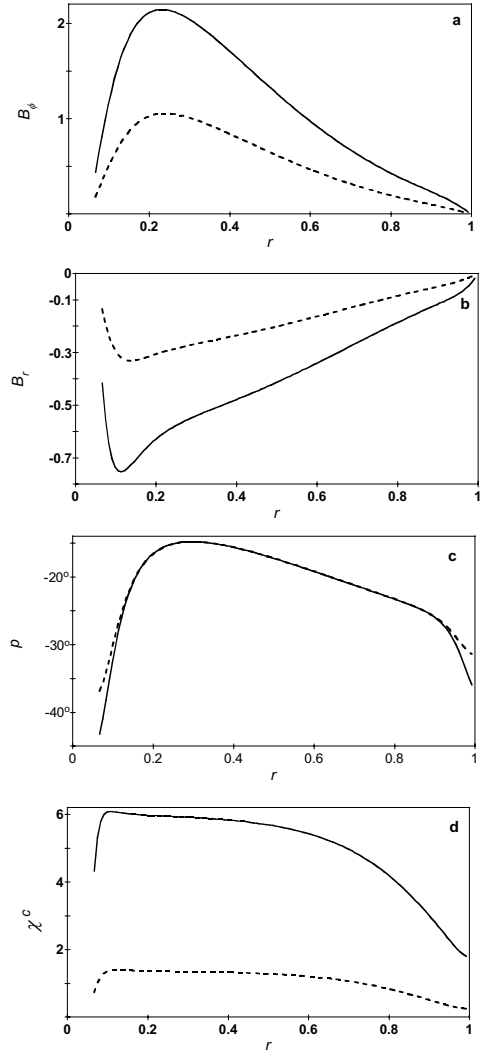}
\caption{Details of the no-$z$ model, with
 dynamo number ${\cal D} = 50$, $h/l=5$, $\lambda = 0.05$,
$|\chi^v(z)|=\sin^2(\pi z/2)$.
a) the toroidal magnetic field $B_{\phi}(r)$,
b) the radial magnetic field $B_r(r)$,
c) the pitch angle $p$,
d) the function $\chi^{c}(z)$.
Broken curves are solutions with $C=3$, solid with $C=10$.}
\label{Fig6.ps}
\end{figure}

Again, we can observe that typically $|\vec{B}|$ is of order
unity. The value $r=1$ can be taken roughly to correspond to 15
kpc in dimensional units, so the radial magnetic field reaches its
maximal value at a galactocentric radius of $3 \dots 4$ kpc, quite
typical for the simplest approximations of galactic rotation
curves. Pitch angles in this radial range are about $-15^\circ
\dots - 20^\circ$, in agreement with available observational data.
The values of $\chi^c$ are in the range $1.5 \dots 6$, so the back
reaction of magnetic field basically modifies the dynamo action.
The main dependencies of the results on the model parameters are
given in Table~1.

We conclude from the the results shown in Table~1 that in the no-$z$ model the
helicity diffusivity $\kappa$
does not play an important role (in contrast to the 1D model of
the previous section).  The only
anomalous thing about solutions with $\kappa = 0$ is an extremely
long (about 50 diffusion times) transient phase when $C = 0.1$.
However this was not a slow growth, but rather a growth over a few
time units followed by a rapid decay and then a slow relaxation. This
feature was not seen when $C = 1,10$, when the magnetic fields grew to
approximately their final strengths over a few time units.

\begin{table}
\caption{Summary of selected calculations for the no-$z$ model.
$E_r/E_\phi$ is the ratio of energies in the radial and azimuthal
magnetic fields, $\langle p \rangle $ and $\langle \chi^c \rangle$
are the averages with respect to radius of pitch angle (in
degrees) and $\chi^c$ in the range $0.20 <r < 0.80$, and
$B_{\phi,m}$ is the maximum value of $B_\phi$.}
\begin{tabular}{|l|l|l|l|l|}
\multicolumn{5}{c}{}\\
\multicolumn{5}{c}{Table 1.}\\
\multicolumn{5}{c}{}\\
\hline
\multicolumn{5}{|c|}{$R_\alpha=2$, $R_\omega=25$, ${\cal D}=50$}\\
\hline
\multicolumn{5}{|c|}{$C=0.1$}\\
\hline
$\kappa$  & $E_r/\bar E_\phi$ & $\langle p \rangle $ & $ \langle
\chi^c \rangle$ & $B_{\phi,m}$ \\
\hline
0   &  0.089 & $-16$ & $-0.67$ & 0.028\\
0.1 &  0.118 & $-18$ &$-0.56$ & 0.060\\
1.0 & 0.113 & $-18$ & $-0.42$ & 0.149\\
10.0 & 0.103 & $-18$ &  $-0.18$ & 0.320\\
\hline
\multicolumn{5}{|c|}{$C=1.0$}\\
\hline
0, 0.1, 1.0, 10.0 & 0.097 & $-18$ & $\approx 0$ & 0.484\\
\hline
\multicolumn{5}{|c|}{$C=10.0$}\\
\hline
0   & 0.104 & $-18$ & 6.22 & 2.19\\
0.1 & 0.103 & $-18$ & 6.1  & 2.19  \\
1.0 & 0.100  & $-18$ & 5.5  & 2.15  \\
10.0 & 0.096 & $-18$ & 2.9 & 0.82  \\
\hline
\multicolumn{5}{|c|}{$C=-1.0$}\\
\hline
1.0 & 0.13 & $-18$ & -0.52 & 0.073\\
10.0 & 0.11 & $-18$ & -0.30 & 0.021 \\
\hline
\multicolumn{5}{|c|}{$R_\alpha = 5$, $R_\omega = 80$, ${\cal D}=400$}\\
\hline
\multicolumn{5}{|c|}{$C=1.0$}\\
\hline
0.1, 1.0, 10.0& 0.01 &$-6$ & $\approx 0$ & 1.93 \\
\hline
\multicolumn{5}{c}{}\\
\end{tabular}
\end{table}

The interaction of the two flux transport coefficients is not
straightforward. For $C=0.1$, field strengths increase with
$\kappa$, whereas for $C=10$, they decrease. When $C=1.0$, the
solutions are remarkably insensitive to the value of $\kappa$,
independent of the value of ${\cal D}$. These comments apply both
to solutions with a modestly supercritical dynamo number (${\cal
D}=50$), and a very supercritical ${\cal D}=400$. The pitch angle
depends predominantly on ${\cal D}$ -- see Table~1.

\section{Discussion and conclusions}

Conventional galactic dynamo models are based usually on a naive
concept of $\alpha$-quenching. From the theoretical point of view
this concept must sooner or later be replaced by a better
elaborated model for the processes of magnetic helicity generation
and transport, which lead to dynamo saturation. In this paper we
have presented a model of such saturation, based as far as
possible physically on first principles rather than on {\it ad
hoc} parameterizations of turbulence. This model is shown to
reproduce successfully  basic features of galactic magnetic fields
and, when used in conjunction with standard galactic dynamo
models, can be used to obtain detailed models of magnetic field
generation in specific galaxies. Perhaps not unexpectedly, we have
to adjust some coefficients of the model in order to obtain this
agreement. However all terms occurring in the model can be
justified from first principles.

Magnetic field distributions obtained with our model are compared
with that from a model with a conventional $\alpha$-quenching and
corresponding values of the  control parameters, in Fig.~7. It can
be seen that results for these models closely resemble each other,
and for the choice $C=3$ they practically coincide; certainly
their deviation is much less than the available accuracy of the
observations. Note that our model, in the case shown in Fig.~7, is
intrinsically quite distant from an $\alpha$-quenched model.
However, for $C=1.0$ and ${\cal D} = -50$ we find $ \langle \chi^c
\rangle \approx 0$, so here the dynamo saturation must be
connected mainly with a conventional $\alpha$-quenching and
$\kappa$ has no influence. We take here $\kappa = 1$ although, as
noted, solutions are insensitive to $\kappa$ for these parameters.
Of course, the last result does not coincide with that for
conventional $\alpha$-quenching, because Eq.~(\ref{D4}) is more
complicated than the conventional $\alpha$-quenching, even when
$\langle \chi^c \rangle \approx 0$.

\begin{figure}
\centering
\includegraphics[width=7cm]{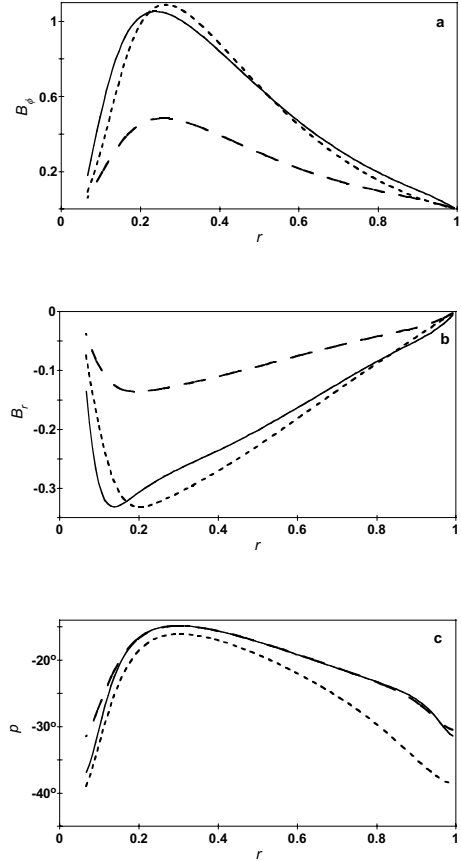}
\caption{Comparison of the equilibrium $r$-profiles for the
no-$z$ model obtained using two kinds of nonlinearities: the
nonlinearity considered here for $C=1$ (dashed) and $C=3$ (solid),
and a standard alpha-quenching $\alpha\propto
1/[1+B^2(\vec{r})/\rho(r)]$ with $\rho=1$ (short-dashed). Notation
is as in Fig.~6.} \label{Fig7.ps}
\end{figure}

From a practical viewpoint, these results lead to a two-fold
conclusion. On one hand the research undertaken can be considered
as a justification of conventional galactic dynamo models based on
a naive $\alpha$-quenching. However the physics underlying our
model is much more complicated. On the other hand it means that
the galactic dynamo, at least in its standard manifestation
considered here, is very robust and the resulting magnetic field
is practically independent of small details of the dynamo model.
It follows that it is necessary to consider more fragile dynamo
regimes in order to isolate features specifically connected with
the details of dynamo saturation mechanisms.

Our analysis has also shown that the detailed form of the flux of
the magnetic helicity is not crucial for the overall state of the
saturated mean magnetic field. The most important factor is to
have a nonzero flux of the magnetic helicity (e.g., in the form of
the transport term or in the form of the diffusive term).

We note that our results clarify the role of magnetic helicity
transport in galactic dynamos in the  mean-field approach only.
However, we do not address the question of the link between the
transport of the total magnetic helicity $ \vec{\cal A} \cdot
\vec{H} $ and the total magnetic field $ \vec{H}$. A recent paper
by Brandenburg et al. (2001b) on the topic of stellar dynamos
using direct numerical simulations isolates many further questions which
are still far from being clear.

Finally, we note that theoretical ideas do not even exclude
negative values of $C$. In this case larger steady state magnetic
fields might be predicted, because the flux transport could then
amplify the mean field and the saturation might be connected with
the diffusive flux only. However the results of simulations with
negative $C$ given in Table~1 demonstrate that the detailed
balance between the various terms in Eq.~(\ref{D4}) is too
complicated for this simple explanation to be valid.

\begin{acknowledgements}
Financial support from NATO under grant PST.CLG 974737, RFBR under
grant 01-02-16158 and INTAS Program Foundation under grant 99-348
is acknowledged. D.~Sokoloff is grateful to a special fund of the
Faculty of Engineering of the Ben-Gurion University of the Negev
for visiting senior scientists. D.~Moss  and D.~Sokoloff thank
the Astronomy Department, University of Uppsala, for its
hospitality whilst the final version of this paper was being
prepared. We are grateful to the anonymous referee for valuable
suggestions.
\end{acknowledgements}

\end{document}